\documentclass[a4paper]{article} 
\usepackage{fullpage}  
\usepackage{amsmath}
\usepackage{amssymb}
\usepackage{graphicx}
\usepackage[font=small, format=hang, labelfont=bf]{caption}
\usepackage[pdfpagelabels, pagebackref, naturalnames]{hyperref}
\usepackage{subfig}

\renewcommand{\figurename}{Fig.}

\iftrue
\usepackage{color}
\hypersetup{%
  pdftitle={}, %
  pdfauthor={}, %
  pdfborder={0 0 1},
  pdfcreator={}, pdfproducer={},
  citebordercolor={0 .667 0},          %
  linkbordercolor={.5812 .0665 .0659}, %
  urlbordercolor={0 0 .667}            %
}
\fi

\author{Hee-Kap Ahn%
  \thanks{Department of Computer Science and Engineering,
    POSTECH, Pohang, Korea. Email:~{heekap@gmail.com}}
  \and Helmut Alt%
  \thanks{Institute of Computer Science, Freie Universit\"at Berlin,
    Germany. Email:~\{alt,christian.knauer\}@inf.fu-berlin.de}
  \and Tetsuo Asano%
  \thanks{School of Information Science, Japan Advanced Institute of
    Science and Technology, Japan. Email:~t-asano@jaist.ac.jp}
  \and Sang Won Bae%
  \thanks{Division of Computer Science, Korea Advanced Institute of
    Science and Technology, Korea.
    Email:~\{swbae,otfried\}@tclab.kaist.ac.kr}
  \and Peter Brass%
  \thanks{Department of Computer Science, City College of New York, USA.
    Email:~{peter@cs.ccny.cuny.edu}}
  \and Otfried Cheong%
  \footnotemark[5]
  \and Christian Knauer%
  \footnotemark[3]
  \and Hyeon-Suk Na%
  \thanks{School of Computing, Soongsil University, Seoul, South
    Korea. Email:~hsnaa@ssu.ac.kr} 
  \and Chan-Su Shin%
  \thanks{School of Electr.~and Inform.~Engineering, Hankuk University
    of Foreign Studies, Yongin, Korea. Email:~cssin@hufs.ac.kr}
  \and Alexander Wolff%
  \thanks{Faculteit Wiskunde en Informatica,
    Technische Universiteit Eindhoven, Eindhoven, the
    Netherlands. WWW: http://www.win.tue.nl/\~{}awolff/}}  

\date{March 8, 2007}

\title{Constructing Optimal Highways\thanks{This research was started
    during the 8th Korean Workshop on Computational Geometry,
    organized by Tetsuo Asano at JAIST, Kanazawa, Japan, Aug.~1--6,
    2005.  O.~Cheong was supported by the Korea Research Foundation.
    C.-S.~Shin was supported by the Hankuk University of Foreign
    Studies Research Fund of 2007.  A.~Wolff was supported by grant
    WO~758/4-2 of the German Research Foundation (DFG).}}

\newtheorem{definition}{Definition}
\newtheorem{theorem}{Theorem}
\newtheorem{lemma}{Lemma}

\newcommand{\R}{\ensuremath{\mathbb{R}}}
\newcommand{\eps}{\varepsilon}
\newcommand{\dopt}{\ensuremath{d_\mathrm{opt}}}
\newcommand{\deltaopt}{\ensuremath{\delta_\mathrm{opt}}}
\newcommand{\deltamed}{\ensuremath{\delta_\mathrm{med}}}

\makeatletter
\newbox\ProofSym
\setbox\ProofSym=\hbox{%
\unitlength=0.18ex%
\begin{picture}(10,10)
\put(0,0){\framebox(9,9){}}
\put(0,3){\framebox(6,6){}}
\end{picture}}
\newenvironment{proof}[1][Proof.]{\O@proof{#1}}{\O@endproof}
\def\O@proof#1{\O@ProofSymtrue\trivlist
   \@topsep\z@\@topsepadd\smallskipamount%
   \@ifstar{\item[]}{\item[\hskip\labelsep\it #1 ]}}
\def\O@endproof{\ifO@ProofSym\relax\hfill\copy\ProofSym\linebreak
  \fi\endtrivlist}
\def\DisplayProofSym{\vskip-\lastskip\vskip-8pt
  \hbox to \hsize{\hfill\copy\ProofSym}\O@ProofSymfalse}
\newif\ifO@ProofSym
\def\QED{\hfill\copy\ProofSym\global\O@ProofSymfalse}

\let\geq\geqslant
\let\leq\leqslant

\makeatletter
\partopsep\z@ \textfloatsep 10pt plus 1pt minus 4pt
\def\section{\@startsection {section}{1}{\z@}{-3.5ex plus -1ex minus
-.2ex}{2.3ex plus .2ex}{\large\bf}}
\def\subsection{\@startsection{subsection}{2}{\z@}{-3.25ex plus -1ex
minus -.2ex}{1.5ex plus .2ex}{\normalsize\bf}}
\def\@fnsymbol#1{\ensuremath{\ifcase#1\or *\or 1\or 2\or 3\or 4\or 5\or
    6\or 7\or 8\or 9\else\@ctrerr\fi}}
\makeatother

\newenvironment{keywords}{\vspace*{2ex}\noindent
  \textit{Keywords:}\hspace{1ex}}{\vspace*{3ex}} 

\begin{document}
\maketitle

\begin{abstract}
  For two points $p$ and $q$ in the plane, a straight line~$h$, called
  a highway, and a real $v>1$, we define the \emph{travel time} (also
  known as the \emph{city distance}) from $p$ and $q$ to be the time
  needed to traverse a quickest path from $p$ to $q$, where the
  distance is measured with speed $v$ on $h$ and with speed $1$ in the
  underlying metric elsewhere.
  
  Given a set $S$ of $n$ points in the plane and a highway speed $v$,
  we consider the problem of finding a \emph{highway} that minimizes
  the maximum travel time over all pairs of points in $S$.  If the
  orientation of the highway is fixed, the optimal highway can be
  computed in linear time, both for the $L_1$- and the Euclidean
  metric as the underlying metric.  If arbitrary orientations are
  allowed, then the optimal highway can be computed in $O(n^{2} \log
  n)$ time. We also consider the problem of computing an optimal pair
  of highways, one being horizontal, one vertical.
\end{abstract}

\section{Introduction}

Many facility location problems are \emph{min-max} problems, where the
task is to place a facility such that the maximum cost incurred by any
customer is minimized.  For example, if the customers are points in
the plane and their cost of using a facility is the Euclidean distance
to the facility, then the center of the smallest enclosing circle is
the optimal facility location that minimizes the maximum distance to
the customers.  Recently, Cardinal and Langerman~\cite{cl-mgflp-06}
introduced the class of \emph{min-max-min} facility location problems,
where customers have the choice of either using or not using the
facility, and their cost is the minimum cost of these two options.
Transport facility location problems are typical min-max-min problems:
a new railway line or highway will not be used by a customer if using
it does not improve the travel time compared to existing means of
transportation.  Cardinal and Langerman consider three such
min-max-min problems, among them the following: Given a set $P$ of
pairs of points, find the highway (a straight line) that minimizes the
maximum travel time over all pairs in~$P$.  Here, the travel time of a
pair of points $(a,b)$ is the time taken to travel from $a$ to $b$,
assuming constant speed anywhere in the plane, infinite speed along
the highway, and travel to and from the highway parallel to the
$y$-axis.  Cardinal and Langerman show how to compute the optimal
highway in expected time linear in the number of pairs.

Previous work considering highways (also called \emph{transportation
  networks} or \emph{roads}) have focused on how to compute quickest
paths among the customers (points) or their Voronoi diagrams under the
metric induced by \emph{given} highways.  Abellanas et
al.~\cite{ahiklmps-pptml1in-01} started work in this area, and
discussed the Voronoi diagram of a point set given a horizontal
highway under the $L_1$-metric.  They later also considered the
problem under the Euclidean metric and studied shortest
paths~\cite{ahiklmps-vdsnh-03}.  Aichholzer et
al.~\cite{aap-qpsscvd-04} introduced the city metric induced by the
$L_1$-metric and a highway network that consists of a number of
axis-parallel line segments.  They gave an efficient algorithm for
constructing the Voronoi diagram and a quickest-path map for a set of
points given the city metric.  The running time of their algorithms
was recently improved by G\"orke et al.~\cite{gsw-ccvdf-07} and Bae et
al.~\cite{bkc-occvd-06}.  Bae et al.~\cite{bc-vdtnep-06} presented
algorithms that compute the Voronoi diagram and shortest paths, using
the Euclidean metric and more general highway networks whose segments
can have arbitrary orientation and speed.  They recently extended
their approach to more general metrics including asymmetric convex
distance functions~\cite{bc-spvdtngp-05}.

In this paper, we, like Cardinal and Langerman, consider the problem
of finding an optimal highway (a straight line) for a given set $S$ of
$n$ points in the plane.  We wish to place a highway such that the
maximum travel time over \emph{all} pairs of points (that is, the
\emph{travel-time diameter} of the point set) is minimized.  Since
there is a quadratic number of pairs, Cardinal and Langerman's
algorithm takes expected quadratic time in this case.  We show how to
make use of the coherence between the pairs of points to get
deterministic near-linear-time algorithms.

We assume that we can travel anywhere in the plane with speed~$1$, and
that we can travel along the highway with a given speed $v > 1$.  We
consider various versions of the problem.  First, we distinguish
whether the highway speed $v$ is infinite (the easiest case) or
finite.  Second, we consider both the $L_{1}$-metric (as in the city
Voronoi diagram) and the $L_{2}$-metric.  In all cases, we show how to
find the optimal highway with a \emph{given} orientation (that is, the
optimal horizontal highway).  In the case of the Euclidean metric, we
also consider how to find the optimal highway if we are free to choose
the orientation.  (We note that chosing the orientation does not make
sense for the Manhattan metric).  Table~\ref{tab:results1} summarizes
our results for all variations of the problem.
\begin{table*}[t]
  \centering
  \begin{tabular}{|c|c|c|c|}
    \hline
    Highway speed & Metric & Fixed orientation & Arbitrary
    orientation\\
    \hline
    Infinite & $L_{1}$ & $O(n)$ & --- \\
    & $L_{2}$ & $O(n)$ & $O(n \log n)$\\
    \hline
    Finite & $L_{1}$ & $O(n)$ & --- \\
    & $L_{2}$ & $O(n)$ & exact: \hfill $O(n^{2} \log n)$ \\
    &         &        & approx.:~~~~~$O(n \log n)$ \\
    \hline
  \end{tabular}
  \caption{Overview of our results for one highway.}
  \label{tab:results1}
\end{table*}

We then consider the problem of placing a \emph{highway cross}, that
is, a pair of a horizontal and vertical highway.  We can determine the
optimal axis-aligned highway cross with infinite speed in $O(n \log
n)$ time, see Section~\ref{sec:cross-infinite}.  For constant speed
the problem becomes considerably harder, even under the $L_1$-metric.
We give an exact $O(n^{4} \alpha(n))$-time algorithm based on
computing minima of upper envelopes.  We also consider approximative
solutions---all our results are summarized in
Table~\ref{tab:results2}.

\begin{table*}[t]
  \centering
  \begin{tabular}{|c|c|c|}
    \hline
    Highway speed & Solution & Running time \\
    \hline
    Infinite & exact & $O(n\log n)$\\
    \hline
    Finite & exact & $O(n^{4} \alpha(n))$\\
    & $(1+\sqrt{2})$-approx. & $O(n \log n)$\\ 
    & $(2+\eps)$-approx.& $O(n^{\color{white}2} \log(1/\eps) \alpha(n) \log n)$\\
    & $(1+\eps)$-approx.& $O(n^{2} \log(1/\eps) \alpha(n) \log n)$\\
    \hline
  \end{tabular}
  \caption{Overview of our results for the highway cross (under the
    $L_1$-metric).} 
  \label{tab:results2}
\end{table*}

Throughout the paper we assume that the input point set $S$ contains
at least three points (if $|S|<3$, it is trivial to find an optimal
highway).

\section{The optimal highway for infinite speed}
\label{sec:highway-infinite}

\begin{figure}
  \centering
  \hfill
  \subfloat[\label{sfl:horizontal}horizontal]%
	   {\includegraphics{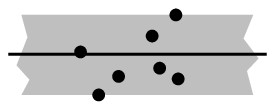}}
  \hfill
  \subfloat[\label{sfl:arbitrary}arbitrary
    orientation]{~~~~\includegraphics{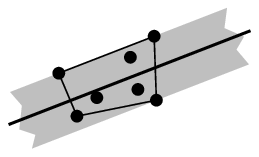}~~~~}
  \hfill\hbox{}
  \caption{Optimal infinite-speed highways (solid lines) and 
    corresponding enclosing figures (shaded).}
  \label{fig:infinite-speed}
\end{figure}

As a warm-up exercise, let us consider the problem of finding the
optimal placement of a horizontal highway, assuming the highway speed
is infinite.
\begin{theorem}
  Given $n$ points in the plane, the middle line of the smallest
  enclosing \emph{horizontal} strip is an optimal \emph{horizontal}
  highway of infinite speed.  It can be computed in linear time.
\end{theorem}
The easy proof is left to the reader.  What is interesting is that the
optimal highway corresponds to a smallest enclosing figure---we will
see this theme repeatedly in the following, see
Figure~\ref{fig:infinite-speed}.
Note that the result holds in any $L_{p}$-metric, as all travel to and
from the highway is parallel to the $y$-axis.

The theorem generalizes to highways of arbitrary orientation in the
Euclidean metric:
\begin{theorem}
  Given $n$ points in the plane, the middle line of the smallest
  enclosing strip is an optimal highway of infinite speed.  It can be
  computed in $O(n \log n)$ time.
\end{theorem}
The algorithm used here is the rotating calipers
algorithm~\cite{t-sgprc-83}.  After computing the convex hull of the
point set, it runs in linear time.

\section{The optimal horizontal highway in the $L_1$-metric}
\label{sec:highway-L1}

Let $\rho \geq 1$ be a real.  We say that a \emph{$\rho$-rhombus} is a
rhombus of aspect ratio~$\rho$.
\begin{theorem}
  \label{lem:l1_main_result}
  Given $n$ points $S$ in the plane, the horizontal axis of symmetry
  of the smallest enclosing axis-aligned $v$-rhombus is an optimal
  horizontal highway for the $L_1$-metric and highway speed~$v$.  It
  can be computed in $O(n)$ time.
\end{theorem}
\begin{proof}
  For a pair of points $p, q \in S$, let $d(p,q) := |x_p - x_q|/v +
  |y_p - y_q|$.  Clearly, $d(p,q)$ is a lower bound for the travel
  time diameter of $S$ for any horizontal highway, and therefore
  $\delta := \max_{p, q\in S}d(p,q)$ is also a lower bound.  We show
  that in fact this bound can be obtained, resulting in an optimal
  highway.
  
  \begin{figure}[bt]
    \centering
    \includegraphics{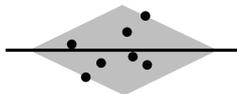}
    \caption{Optimal constant-speed horizontal highway (solid
      lines) and  corresponding enclosing figure (shaded).}
    \label{fig:constant-speed}
  \end{figure}
  We observe that the point set can be enclosed in a rhombus with
  horizontal diagonal $\delta v$ and vertical diagonal~$\delta$.  For
  an example with $v=2$, see Figure~\ref{fig:constant-speed}.  If we
  place a horizontal highway along the horizontal diagonal of this
  rhombus, then any point in the rhombus has travel time at most
  $\delta/2$ to the center of the rhombus.  This implies that the
  travel-time diameter is at most $\delta$.

  The computation boils down to computing minimum and maximum $y$-axis
  intercepts among all lines through points of $S$ of slope $1/v$
  and~$-1/v$.
\end{proof}

\section{The optimal highway in the Euclidean Metric} 
\label{sec:highway-L2}

In this section we consider the Euclidean metric in the plane, and a
highway of finite speed~$v > 1$.

\begin{theorem}
  \label{thm:l2-horizontal}
  Given $n$ points $S$ in the plane, an optimal horizontal speed-$v$ highway
  under the Euclidean metric can be computed in $O(n)$ time.
\end{theorem}

\begin{figure*}
  \subfloat[\label{fig:pq-eucl}Quickest \mbox{$p$--$q$
  path.}]{\includegraphics{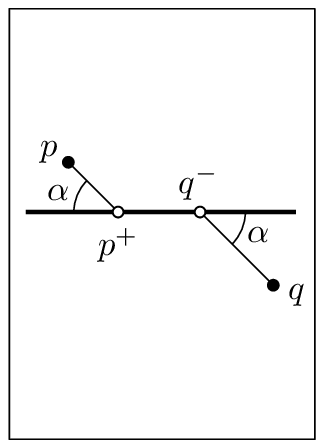}} 
  \hfill
  \subfloat[\label{fig:pq-negative-a}The line $pq'$ forms an angle 
  greater $\alpha$ with the $x$-axis.]{\includegraphics{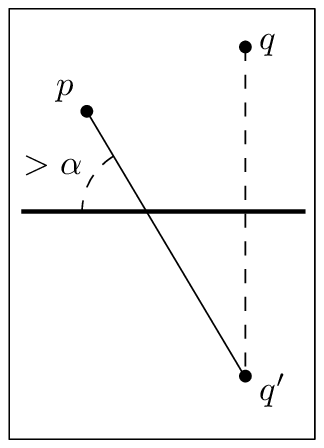}}
  \hfill
  \subfloat[\label{fig:pq-negative-b}The $\eta$-distance counts the
  time on the highway \emph{negative}.]{~\includegraphics{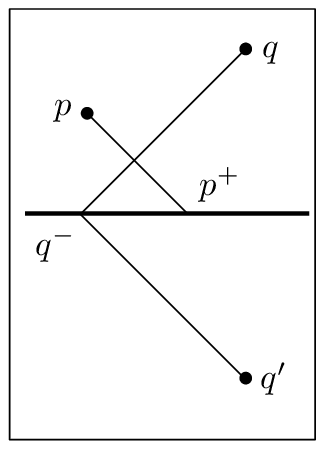}~}
  \hfill
  \subfloat[\label{fig:pq-negative-c}The path from $p$ to $q$
    self-intersects in $x$.]{\includegraphics{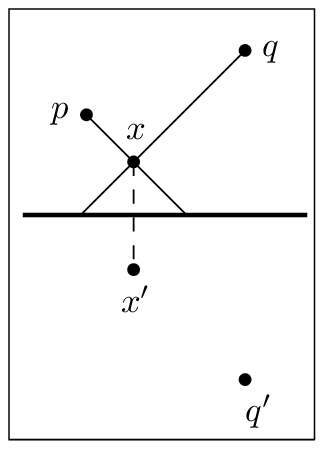}}

  \caption{Optimal highway under the $L_2$-metric.}
  \label{fig:pq-negative}
\end{figure*}

\begin{figure}[bt]
  \centering
  \includegraphics{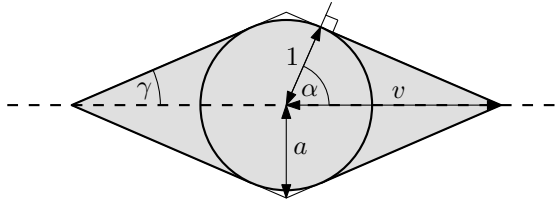}
  \caption{Unit disk in the travel-time metric (shaded) and smallest
    enclosing $(\tan\alpha)$-rhombus.}
  \label{fig:eucl-unit-disk}
\end{figure}

\begin{proof}
  The quickest path (that is, the path with shortest travel-time)
  between two points $p$ and $q$ has one of two
  forms~\cite{ahiklmps-vdsnh-03}: It is either the segment $pq$; or a
  path consisting of three segments $pp^{+}$, $p^{+}q^{-}$, $q^{-}q$,
  where $p^{+}$ and $q^{-}$ are points on the highway, and the lines
  $pp^{+}$ and $qq^{-}$ form an angle of $\alpha=\arccos 1/v$ with the
  highway, see Figure~\ref{fig:pq-eucl}.

  Now let us define a norm $\eta(x,y)$ on $\R^{2}$ as
  \[
  \eta(x,y) = |x|\cos \alpha + |y| \sin \alpha.
  \]
  Since $0 < \alpha < \pi/2$, we have $\eta(x,y) > 0$ unless $(x,y) =
  (0,0)$, and $\eta$ is indeed a norm.

  Let $p$ and $q$ be two points such that the highway is inbetween $p$
  and $q$ and such that the shortest path between $p$ and $q$ makes
  use of the highway.  Then the travel time from $p$ to $q$ is
  $\eta(q-p)$.  Indeed, let $(x, y) = q - p$, and assume $x, y \geq
  0$. Then the travel time from $p$ to $q$ is $ay + (x - by)/v$, where
  $a = 1/\sin \alpha$ and $b = 1/\tan \alpha = a/v$.  Using $1 -
  1/v^{2} = 1/a^{2}$, we get $ay + (x - by)/v = \eta(x,y)$.

  When the highway cannot be used because the line $pq$ forms an angle
  larger than $\alpha$ with the highway, then the travel time is
  simply the Euclidean distance $d(p,q)$, and $\eta(q-p)$ is an
  underestimate.

  The unit circle under the norm~$\eta$ is a $(\sin\alpha /
  \!\cos\alpha)$-rhombus, i.e., a $(\tan \alpha)$-rhombus.  We can
  find the smallest such rhombus enclosing a given set $S$ of $n$
  points in linear time.  This means we have the smallest factor
  $\delta > 0$ such that the $S$ fits in the rhombus $R$ with corners
  $(0, \delta a)$, $(0, -\delta a)$, $\delta v, 0)$, and $(-\delta v,
  0)$ (after translating the point set).

  We claim that the $x$-axis is now an optimal highway.  We already
  know that there is a pair of points whose $\eta$-distance is
  $2\delta$, so this is a lower bound on the diameter.  We now show
  that for any pair of points, either their travel time (with respect
  to the highway at $y = 0$) is at most $2\delta$, or they cannot use
  \emph{any} horizontal highway.

  For any two points $p, q$ in the rhombus $R$, we have $\eta(q-p)
  \leq 2\delta$.  This means that if the highway lies inbetween, then
  we are already done.  So assume that both $p$ and $q$ lie above the
  highway, and such that they \emph{can} use a horizontal highway.
  Let $q'$ be the reflection of $q$ around the highway.  Since $R$ is
  symmetric with respect to $y = 0$, $q'$ is also in $R$, and if $p,
  q'$ can use a horizontal highway, then their travel-time distance is
  at most $2\delta$, implying that the travel time from $p$ to $q$ is
  also at most $2\delta$.

  It remains to consider the case that $p,q'$ cannot use a horizontal
  highway.  This means that the line $pq'$ forms an angle larger than
  $\alpha$ with the $x$-axis, see Figure~\ref{fig:pq-negative-a}.
  Note that $\eta(q'-p)$ still has a geometric meaning: There is a
  path from $p$ to the highway, then \emph{backwards} along the
  highway, then straight to $q$, see Figure~\ref{fig:pq-negative-b}.

  The $\eta$-distance measures the whole travel time, but counting the
  time on the highway \emph{negative}.  Reflecting the last segment of
  this path back around the highway, we obtain a path from $p$ to $q$
  with travel time $\eta(q'-p)$, still counting time spent on the
  highway negative.  But now observe that this path self-intersects in
  a point~$x$, see Figure~\ref{fig:pq-negative-c}.  Let $x'$ be the
  reflection of $x$.  Then $\eta(q'-p) = d(p,x) + \eta(x'-x) +
  d(x',q') = d(p,x) + \eta(x'-x) + d(x,q) \geq d(p,x) + d(x, q) \geq
  d(p,q)$ (using $\eta \geq 0$).  It follows $d(p,q) \leq \eta(q'-p)
  \leq 2\delta$.
\end{proof}
Again we found an optimal highway by computing a minimal enclosing
shape.  Interestingly, the shape to be minimized is not the unit
circle under the travel-time metric.  If the highway is the $x$-axis
then the unit circle in the travel-time metric is the convex hull of
the points $(0,v)$ and $(0,-v)$ and the Euclidean unit circle centered
at the origin, see the shaded region in Fig.~\ref{fig:eucl-unit-disk}.
The reader may wish to verify that finding the smallest copy of this
(unit-disk) shape enclosing $S$ does not give the optimal horizontal
highway.  The rhombus we are minimizing instead is the
``rhombus approximation'' of this shape, see
Fig.~\ref{fig:eucl-unit-disk}.

The very simple linear-time algorithm above results in a horizontal
highway that minimizes the travel-time diameter of the point set, but
it does not actually tell us what the travel-time diameter is.
Surprisingly, it is actually not possible to compute the travel-time
diameter within the same time bound, as the following lemma shows.
\begin{lemma}
  In the algebraic decision-tree model, computing the travel-time
  diameter for a set of $n$ points and a given highway takes $\Omega(n
  \log n)$ time.
\end{lemma}

\begin{proof}
  The following problem has a lower bound of $\Omega(n \log n)$ in the
  algebraic decision-tree model: Given two sets $A$ and $B$ of $n$
  real numbers, is $A \cap B = \emptyset$?  We show how to transform
  this problem in linear time into a decision instance of the diameter
  problem.
 
  Our instance consists of a set $A'$ of $n$ points and a set $B'$ of
  $n$ points, computed from $A$ and $B$.  All points lie on the unit
  circle.  We first scale all numbers in $A$ and $B$ so that they are
  close to zero (depending on $v$).  For each $a$ in $A$, we create
  the point $(a,\sqrt{1 - a^2})$.  For each $b$ in $B$, we create the
  point $(-b,-\sqrt{1-b^2})$.  Note that since the points are close to
  the $y$-axis, no horizontal highway can be used to speed up the
  connection between $A'$ and $B'$, and so the diameter of the set is
  simply the Euclidean diameter. It follows that the diameter of $A'
  \cup B'$ is $2$ if and only if $A$ and $B$ contain a common number.
\end{proof}

\begin{theorem} 
  \label{thm:highway-euclid}
  Given a set $S$ of $n$ points in the plane, the optimal highway with
  speed $v$ can be found in $O(n^{2} \log n)$ time.
\end{theorem}
\begin{proof}
  First we compute the convex hull $C$ of the point set~$S$.  Then we
  use the rotating-calipers algorithm~\cite{t-sgprc-83} to compute the
  function $w: [0,\pi) \to \R^+$ that maps an angle $\phi$ to the
  width of the smallest strip that contains $C$ and makes angle~$\phi$
  with the positive $x$-axis.  The function~$w$ consists of at most
  $n$~pieces, each of which is a trigonometric function that can be
  computed explicitly in constant time.  Let $\gamma=\pi/2-\alpha$ be
  the angle formed by the main diagonal and the sides of a $(\tan
  \alpha)$-rhombus (where $\cos \alpha = 1/v$), see
  Fig.~\ref{fig:eucl-unit-disk}.  Then the function $u(\phi) = \max \{
  w(\phi - \gamma \bmod \pi), w(\phi+\gamma \bmod \pi) \}$ maps~$\phi$
  to the width of a smallest $(\tan \alpha)$-rhombus that contains $C$
  and whose main diagonal forms an angle of~$\phi$ with the positive
  $x$-axis.
  
  As we saw above in the proof of Theorem~\ref{thm:l2-horizontal}, the
  main diagonal of this rhombus is an optimal highway for this
  orientation, and the travel-time diameter $z(\phi)$ for this highway
  is
  \[
  z(\phi) = \max \{ u(\phi), \max_{p,q} d(p,q) \},
  \]
  where the maximum is taken over all pairs $p,q \in S$ that cannot
  use any highway with orientation~$\phi$.  This is the case if the
  vector $p-q$ makes an angle larger than $\alpha$ with the highway
  orientation.  We define now, for every pair $p,q \in S$, a function
  $f_{pq}: [0,\pi) \rightarrow \R^+_0$ with $f_{pq}(\phi)=d(p,q)$ if
  the angle between the vector $p-q$ and the positive $x$-axis is
  inbetween $\phi + \pi/2 - \alpha$ and $\phi + \pi/2 + \alpha$
  modulo~$\pi$, and $f_{pq}(\phi)=0$ otherwise. We can then rewrite
  the optimal travel-time diameter for orientation $\phi$ as
  \[
  z(\phi) = \max \{ u(\phi), \max_{p,q \in S} f_{pq}(\phi) \}.
  \]
  The graph of $z$ is the upper envelope of $n \choose 2$ horizontal
  segments and of the graph of $u$ (which has $O(n)$ breakpoints).
  Thus the graph of $z$ can be computed in time $O(n^{2} \log n)$, and
  we find an optimal orientation by picking a lowest point on this
  graph.
\end{proof}

The function $u$ used in the proof here can be computed in $O(n \log
n)$ time, and within the same time bound we can pick an orientation
$\phi$ for which $u$ is minimal.  This means that $u(\phi)$ is a lower
bound for the travel-time diameter of \emph{any} highway for $S$.  Let
$R$ be the smallest enclosing $(\tan \alpha)$-rhombus with
orientation $\phi$.  Then any point in $R$ has travel-time distance at
most $u(\phi)/(2\sin\alpha)$ from the center of $R$, see
Fig.~\ref{fig:eucl-unit-disk}.  This implies that the highway
$h_\mathrm{app}$ through the main diagonal of $R$ has a travel-time
diameter of at most $u(\phi)/\sin\alpha$, which is within a factor of
$1/\sin\alpha = \sqrt{v^2/(v^2-1)}$ from optimal.  On the other hand,
any highway yields at least the same (factor-$v$) approximation as
building no highway at all.  It is easy to see that the maximum of the
function $\min \{v, \sqrt{v^2/(v^2-1)}\}$ for $v \geq 1$ is attained at
$v=\sqrt{2}$ and has a value of $\sqrt{2}$.  Thus $h_\mathrm{app}$ is
in fact at least a $\sqrt{2}$-approximation of the optimal highway.

\begin{theorem} 
  \label{thm:highway-euclid-approx}
  Given $n$ points $S$ in the plane, a speed-$v$ highway with
  travel-time diameter within a factor of $\min \{v,
  \sqrt{v^2/(v^2-1)}\} \leq \sqrt{2}$ from optimal can be found in $O(n
  \log n)$ time.
\end{theorem}

The approximation factor of this algorithm depends on~$v$, and since
\[
\sqrt{\frac{v^2}{v^2-1}} = \sqrt{1 + \frac{1}{v^2-1}} \leq 1 +
\frac{1}{2(v^2-1)},
\] 
the factor tends to~$1$ with growing~$v$ very quickly.  For instance, 
for speeds $v = 2$, 3, and 10, the factors are at most 1.16,
1.06, and 1.005, respectively.  

\section{The optimal highway cross for infinite speed} 
\label{sec:cross-infinite}

Now we consider the problem of placing more than one highway.  Observe
that multiple parallel highways with the same speed do not reduce the
maximum travel time because the quickest path using several highways
can be simulated with only one highway.  Instead we investigate
highway crosses, that is, pairs of highways that intersect
perpendicularly.  We give algorithms for computing the optimal
\emph{axis-aligned} highway cross.

\begin{definition}
  An \emph{enclosing cross} for a point set $S$ is the union of a
  horizontal and a vertical strip of equal width containing $S$.
\end{definition}

\begin{lemma} \label{lem:cross-infinite}
  The travel-time diameter of the optimal axis-aligned highway cross
  with infinite speed equals the width of the smallest enclosing
  cross.
\end{lemma}
\begin{proof}
  Let $\delta$ be the travel-time diameter, and let $\delta'$ be the
  width of a smallest enclosing cross~$C$.

  We first show that $\delta' \leq \delta$: Let $h_{1}, h_{2}$ be a
  pair of optimal highways.  We assign each point in $S$ to its
  closest highway so that $S$ is partitioned into two subsets: one
  consisting of points closer to the horizontal highway and the other
  consisting of points closer to vertical highway. We put around each
  highway the narrowest strip containing all the points assigned to
  the highway.  Then both strips have width at most $\delta$,
  otherwise there are two points in the wider strip whose travel-time
  distance is larger than $\delta$.  Therefore we can obtain an
  enclosing cross of width $\delta$ by widening each strip until its
  width becomes $\delta$.  Since $\delta'$ was minimal, we have
  $\delta' \leq \delta$.

  It remains to show $\delta \leq \delta'$: We place highways in the
  middle of each strip of $C$. This results in a pair of highways with
  travel-time diameter at most $\delta'$.  Since $\delta$ is optimal,
  we have $\delta \leq \delta'$.
\end{proof}

Note that once again the optimal facility corresponds to a minimal
enclosing shape.  This shape can be computed efficiently.

\begin{theorem} \label{thm:cross-infinite}
  Given $n$ points in the plane, the optimal axis-aligned highway
  cross for infinite speed corresponds to the smallest enclosing strip
  cross.  It can be computed in $O(n \log n)$ time.
\end{theorem}

\begin{proof}
  The characterization follows from Lemma~\ref{lem:cross-infinite}.
  The smallest enclosing cross of a set $S$ of $n$ points can be found
  as follows.
\begin{enumerate}
\item \label{enum:presort} %
  We presort the points by their $x$- and by their $y$-coordinates.
\item \label{enum:decide} %
  For a given width $\omega > 0$, we can decide in linear time whether
  an enclosing cross of width $\omega$ exists. If it is the case, the
  enclosing cross can be found in the same time.  Our decision
  algorithm is as follows.  We slide a vertical strip $V$ of width
  $\omega$ across the point set from left to right.  We maintain a
  horizontal strip $H$ of smallest width containing all the points not
  in $V$.  For each point entering $V$ from the right or leaving $V$
  from the left, we update $H$ accordingly. If the width of $H$
  ever becomes $\omega$ or less, we answer ``yes'' and report an
  enclosing cross. Otherwise, we answer ``no''.
\item \label{enum:list} %
  The width of the smallest enclosing cross is in the list of numbers
  $L = L_x \cup L_y$, where $L_x = \{x_j-x_i \mid 1\leq i < j\leq n\}$
  and $x_{1} \leq x_{2} \leq \cdots \leq x_{n}$ is the sorted sequence of
  $x$-coordinates.  The list $L_y$ is defined analogously based on the
  sorted sequence of $y$-coordinates.
\item \label{enum:binary} %
  Consider the matrix $A$ with $A[i,j]=x_j-x_{n-i+1}$.  The rows and
  columns of $A$ are sorted in ascending order. Using the technique of
  \cite{fj-gsrsm-84}, we can determine the $k$-th element of
  such a \emph{sorted matrix} in $O(n)$ time without constructing $A$
  explicitly.  This gives us a way to do binary search on $L_x$.  This
  search consists of $O(\log n)$ steps, each of which first invokes
  the algorithm of Frederickson and Johnson to find the median of the
  remaining elements in $L$ and then calls the decision algorithm of
  stage~\ref{enum:decide}.  Thus, the total runtime is $O(n\log n)$.
  Likewise we can search for the smallest value in $L_y$ for which an
  enclosing cross exists.  Finally we return the minimum of the two
  values.  Notice that the algorithm in stage~\ref{enum:decide}
  computes not only the width of the smallest enclosing cross but also
  the cross itself. \QED
\end{enumerate}
\end{proof}

\section{The optimal axis-aligned highway cross for finite speed}
\label{sec:cross-L1}

In this problem we have been unable to characterize the optimal
solution by a smallest enclosing shape.  There is no natural
``center'' for the problem: sometimes there are critical paths where a
point connects to the highway that is \emph{further} away.

\begin{theorem}
  \label{thm:cross-exact}
  The optimal axis-aligned speed-$v$ highway cross under the
  $L_1$-metric can be computed in $O(n^{4}\alpha(n))$ time.
\end{theorem}

\begin{proof}
  The optimal solution corresponds to the lowest point on the upper
  envelope of the pairwise distance functions, see
  Figure~\ref{fig:pq-cross}.  Since these functions are piecewise
  linear and of constant complexity, their upper envelope can be
  computed in $O(n^{4}\alpha(n))$ time~\cite{egs-ueplf-89}.
\end{proof}

\subsection{The decision problem} \label{sub:decision}

We now present an algorithm that decides for a given $\delta > 0$
whether there is an axis-aligned speed-$v$ highway cross such that the
resulting travel-time diameter is at most $\delta$.  This will be used
as a subroutine for finding approximations of the optimal highway
cross, see Section~\ref{sec:approx-cross}.

\begin{theorem} \label{thm:decision}
  Given a set $P$ of $m$ pairs of points in the plane, a speed $v >
  1$, and a parameter $\delta > 0$, we can decide in time
  $O(m\alpha(m)\log m)$ whether there is an axis-aligned highway cross
  of speed~$v$ such that the travel-time distance of all pairs in $P$
  is at most~$\delta$.  If the answer is positive, we can compute such
  a highway cross within the same time bound.
\end{theorem}

\begin{proof}
  For points $\sigma,p,q \in \R^{2}$, let $d_{\sigma}(p,q)$ denote the
  travel-time distance between $p$ and $q$, assuming an axis-aligned
  highway cross with speed~$v$ has been placed (with center)
  at~$\sigma$.
  We define the region 
  \[
  R(p,q) := \{ \sigma \in \R^{2}\mid d_{\sigma}(p,q) \leq \delta \}.
  \]
  We observe that the answer to the decision problem is positive if
  and only if $\bigcap_{(p,q) \in P}R(p,q)$ is not empty.

  The shape of the region $R(p,q)$ depends on $\delta$. Let $w, h$ be
  the horizontal and vertical distance of points $p$ and $q$.  If
  $\delta < (w+h)/v$, then $R(p,q)$ is empty.  If $(w+h)/v \leq \delta
  < \min\{w + h/v, h + w/v\}$, then $R(p,q)$ consists of two convex
  quadrilaterals. If $\min\{w + h/v, h + w/v\} \leq \delta
  < \max\{w + h/v, h + w/v\}$, then $R(p,q)$ is infinite in one
  (axis-parallel) direction.  If $\max\{w + h/v, h + w/v\} \leq \delta
  < w + h$, then $R(p,q)$ is infinite in both axis-parallel
  directions. Finally, if $w+h \leq \delta$, then $R(p,q) = \R^{2}$.
  See Figure~\ref{fig:pq-4regions}.

  \begin{figure}
    \centering
    \includegraphics{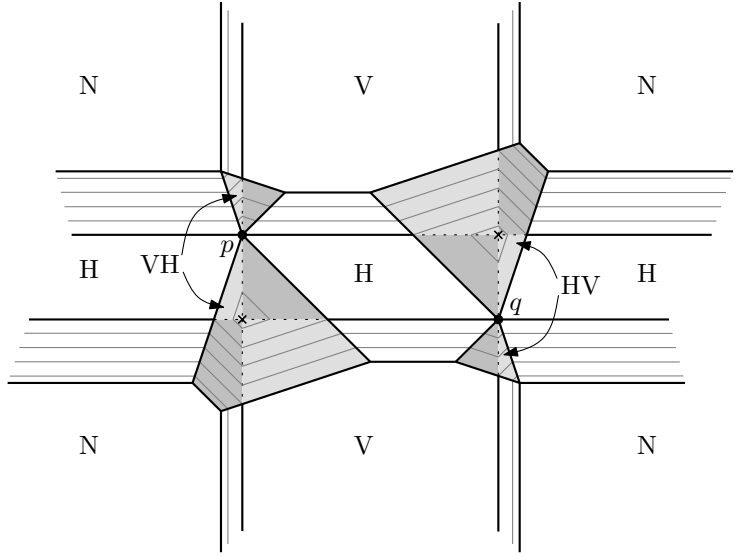}
    \caption{Travel-time distance between two points~$p$ and~$q$ for
      an axis-aligned highway cross centered at $(x,y)$.  The graph of
      this function has $31$ faces of $15$ different orientations. The
      two points marked $\times$ are the lowest, that is, those where
      the corresponding highway crosses minimize the travel time
      from~$p$ to~$q$.  The thin dark gray lines are contour lines.
      If a highway cross is centered in a V- or H-region, the vertical
      and horizontal highway, respectively, is used by a quickest
      $p$--$q$ path.  In the VH- and HV-region both highways are used
      in the corresponding order.  In the N-regions no highway is
      used.}
    \label{fig:pq-cross}
  \end{figure}

  \begin{figure*}[tbp]
    \centering
    \includegraphics{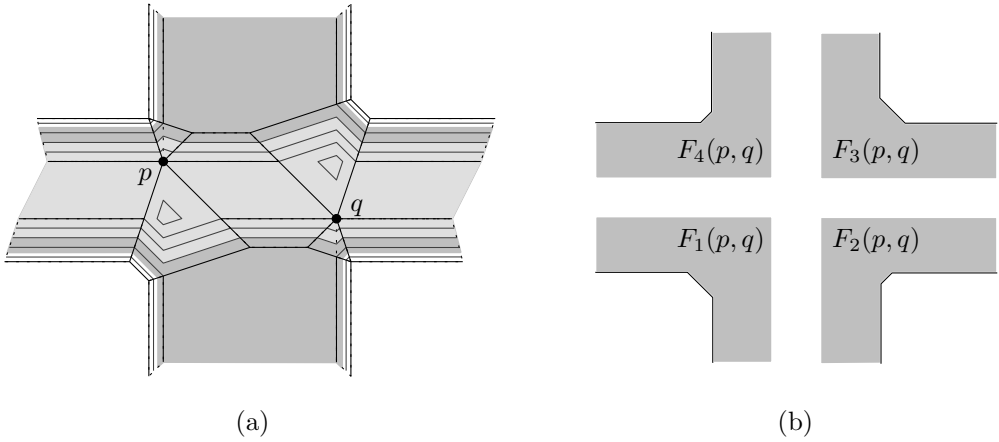}
    \caption{(a) the regions $R(p,q)$ when $\min\{w + h/v, h + w/v\}
      \leq \delta < \max\{w + h/v, h + w/v\}$ (light gray region) and when
      $\max\{w + h/v, h + w/v\} \leq \delta < w + h$ (dark and light
      gray regions). (b) the dark and light gray regions can be expressed 
      as the intersection of the four types of regions $F_1, F_2, F_3$,
      and $F_4$.} 
    \label{fig:pq-4regions}
  \end{figure*}

  Let us call a planar region $F$ $(a,b)$-monotone if for every point
  $(x,y) \in F$ and any $\lambda \geq 0$ the point $(x + \lambda a, y
  + \lambda b)$ is also in~$F$.  We observe that $R(p,q)$ can be
  expressed as the intersection of four regions $F_{i}(p,q)$, $i =
  1,2,3,4$, where $F_{1}(p,q)$ is $(1,1)$-monotone, $F_{2}(p,q)$ is
  $(-1,1)$-monotone, $F_{3}(p,q)$ is $(-1,-1)$-monotone, and
  $F_{4}(p,q)$ is $(1,-1)$-monotone.  Figure~\ref{fig:pq-4regions}
  shows an example of $R(p,q)$, which can be expressed as the
  intersection of $F_1, F_2, F_3$, and $F_4$.  Each region is bounded
  by a polygonal curve of constant complexity.  $F_{i} :=
  \bigcap_{(p,q)\in P}F_{i}(p,q)$ is the lower envelope of a set of
  $O(m)$ line segments, has complexity
  $O(m\alpha(m))$~\cite{sa-dssga-95}, and can be computed in time $O(m
  \log m)$.  The intersections $F_{1} \cap F_{3}$ and $F_{2} \cap
  F_{4}$ can be computed by a plane sweep in time $O(m \alpha(m) \log
  m)$.  We are left with two regions of complexity $O(m\alpha(m))$,
  and we need to determine whether their intersection is empty.  While
  we do not know how to bound the complexity of this region, we can
  test emptiness in $O(m\alpha(m) \log m)$ time, by a simple plane
  sweep that stops as soon as a point in the intersection is found.
  Since any intersection between edges of the two regions implies that
  the intersection is not empty, this runs in the claimed time bound.
  If such an intersection is found, it is a center for highway cross
  with travel-time diameter at most~$\sigma$.
\end{proof}

\subsection{Further observations}

Suppose we could characterize the travel-time diameter given the
optimal highway cross to get a compact list $L$ of candidate values as
in the case of the infinite-speed highway cross.  Then we could do
binary search on $L$ using the decision algorithm of
Theorem~\ref{thm:decision}.

Given the travel-time graph $\Gamma_{pq}$ for each pair of points $p$
and $q$ (see Figure~\ref{fig:pq-cross}), consider the upper envelope
over all these graphs as in Theorem~\ref{thm:cross-exact}.  The
minimum occurs at a vertex of this upper envelope.  Thus there are
always at most three pairs---that is, at most six points---that define
this vertex.  Locally, in a neighborhood of the optimal highway, these
at most six points alone have the same optimal highway cross.
However, the lower envelope for these three pairs might have other
minima, and the original solution might not be the global minimum.  We
do not know whether this is the case, and whether it could perhaps be
resolved by considering a few more pairs.  If that was possible, than
we could apply Chan's technique~\cite{c-garot-99} (as Cardinal and
Langerman do~\cite{cl-mgflp-06}), in order to get a randomized
algorithm for the optimization problem whose expected running time is
asymptotically the same as the running time of the decision algorithm.

\begin{figure}
  \centering
  \includegraphics{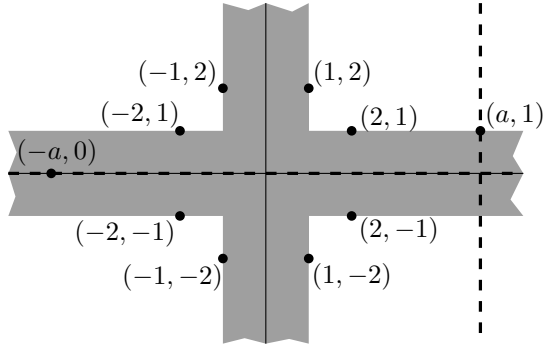}
  \caption{Here the optimum speed-$v$ highway cross (dashed) is not
    contained in the optimum speed-$\infty$ strip cross (shaded).}
  \label{fig:different-strip}
\end{figure}

The optimum axis-aligned highway cross for finite speed need not be
contained in the strip cross for infinite speed, see
Figure~\ref{fig:different-strip}: take the points $(-2,1)$, $(-1,2)$,
$(1,2)$, $(2,1)$, $(2,-1)$, $(1,-2)$, $(-1,-2)$, $(-2,-1)$---that is, an
octagon, contained in the strip cross $([-1,1] \times \R) \cup (\R
\times [-1,1])$)---plus the two points $(-a, 0)$ and $(a,1)$ for
$a>3v$.  Then the coordinate axes are the optimum highway cross for
infinite speed.  It has travel-time diameter~$2$, so the above two
strips are the optimum cover.  But for any finite speed $v>1$, the
highway cross $x=a$ and $y=0$ yields a diameter of $(2a+1)/v$, which
is better than the diameter $(2a/v)+1$ caused by the coordinate axes
being highways.

\begin{figure}
  \centering
  \includegraphics{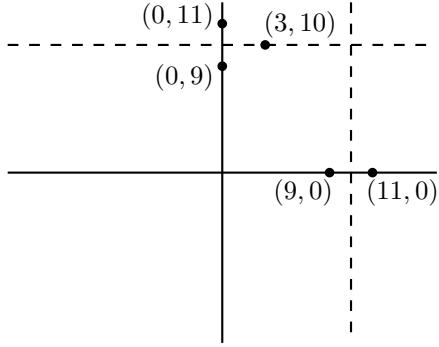}
  \caption{After adding the point $(3,10)$ the optimal speed-$\infty$
    highway cross changes (from solid to dashed), but the new point
    does not occur in any diametral pair.}
  \label{fig:incremental-change}
\end{figure}

The following argument also rules out any simple incremental
algorithm: Even for infinite speed there are point sets such that the
addition of one point changes the diameter, and the new point does not
occur in any diametral pair.  An example (see
Figure~\ref{fig:incremental-change}) for infinite speed is given by
the points $(0,9)$, $(0,11)$, $(9,0)$, $(11,0)$ since now the
coordinate axes are an optimal highway cross---with a diameter of~$0$.
If we add the point $(3,10)$, the optimal highway cross is centered at
$(10,10)$ and has a diameter of~$2$.

\section{Approximations for the optimal axis-aligned highway cross}
\label{sec:approx-cross}

Given a set $S$ of $n$ points.  Let $C$ be the smallest enclosing
cross for $S$, and let $h_{1}, h_{2}$ be the middle line of each strip
of $C$.  We call $h_{1}, h_{2}$ the \emph{median highways} for~$S$.

\begin{lemma} \label{lem:median} %
  The travel-time diameter \deltamed\ of the median highways (with
  speed~$v$) is at most $2 + 1/v$ times the travel-time diameter
  \deltaopt\ of an optimal axis-aligned speed-$v$ highway cross for
  $S$.  There are point sets $S$ where for $v\geq\sqrt{3}$ the
  travel-time diameter of the median highways is at least $2 -
  1/(v+2)$ times the optimum.
\end{lemma}

\begin{figure}[tbp]
  \centering
  \includegraphics{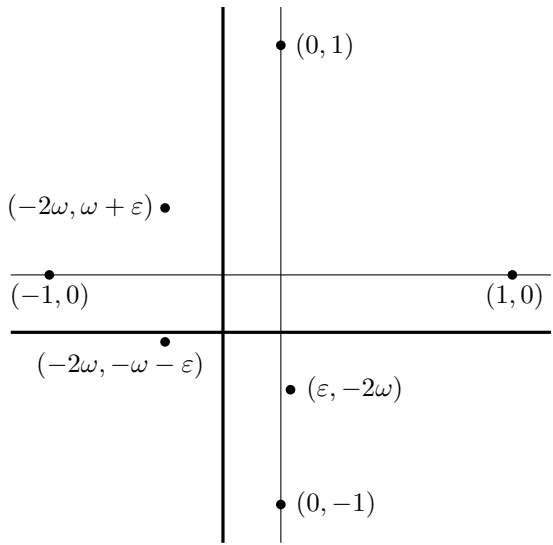}
  \caption{The median highway for $v=2$.}
  \label{fig:median}
\end{figure}

\begin{proof}
  We can scale $S$ such that its $L_{1}$-diameter is~$2$---this does
  not change the travel-time ratio.  Let $w$ be the width of $C$ after
  scaling.  Observe that $\deltaopt \geq 2/v$, as there are points at
  $L_{1}$-distance~$2$.  Furthermore, we have $\deltaopt \geq w$,
  since using the optimal highways at infinite speed cannot achieve
  diameter less than $w$.

  On the other hand, $\deltamed \leq w + (2+w)/v$, since any
  point can reach a point on the highways at distance at most
  $w/2$, and the maximum distance of such points on the highways
  is at most $2 + w$.  This implies $\deltamed \leq (1+1/v) w +
  2/v \leq (1+1/v)\deltaopt + \deltaopt = (2 + 1/v) \deltaopt$.
  
  For the lower bound example, let speed $v > 1$ be given, 
  and set parameter $\omega = 1/(v+2)$. We will construct a point set $S$
  such that the smallest enclosing cross has width $2\omega$, the
  median highway has travel-time diameter $4/v - 2/(v(v+2))$, and the
  optimal highway cross has travel-time diameter at most $2/v$,
  implying the lower bound.

  Let $\eps > 0$ be very small.  Our point set $S$ consists of the
  points $(1,0)$, $(-1, 0)$, $(0,1)$, $(0,-1)$, $(-2\omega, \omega +
  \eps)$, $(-2\omega, -\omega - \eps)$, $(\eps, -2\omega)$, as in
  Figure~\ref{fig:median}.  We claim that $S$ has a unique smallest
  enclosing strip of width $2\omega$, centered around the lines $x =
  -\omega$ and $y = -\omega$.  Indeed, the line $x = 0$ must be in the
  vertical strip (otherwise the horizontal strip would have width at
  least~$2$), while the line $y= 0$ must be in the horizontal strip.
  Similarly, the line $x = -2\omega$ must be in the vertical strip as
  well, and this now fixes the vertical strip of width $2\omega$
  around the line $x = -\omega$.  It follows that the remaining point
  $(\eps, -2\omega)$ is in the horizontal strip, fixing that strip
  around $y = -\omega$.

  The median highway cross has travel-time diameter $2\omega + (2 +
  2\omega)/v = 4/v - 2/(v(v+2))$ if the diameter is less than or equal
  to the $L_1$ distance of these points (note that the diameter is
  determined by $(1,0)$ and $(0,1)$.) That is, for $v\geq\sqrt{3}$ the
  median highway cross has travel-time diameter $4/v - 2/(v(v+2))$.
  Consider now a highway cross with center at the origin.  The four
  outer points and the point $(\eps, -2\omega)$ can be reached from
  the origin within travel time~$1/v$.  The remaining two points can
  be reached with travel time $\omega (1+2/v)$ (ignoring all
  $\eps$-terms).  The travel-time distance between these two points is
  $2\omega$, and so the travel-time diameter is bounded by
  \[
  \max\{ 2/v, 1/v + \omega(1+2/v), 2\omega \} = 2/v. 
  \]
  \DisplayProofSym
\end{proof}

We can improve the result in Lemma~\ref{lem:median} by a simple
observation.

\begin{theorem} \label{thm:constbootstrap} %
  Given a set $S$ of $n$ points we can compute in $O(n \log n)$ time
  an axis-aligned highway cross whose travel-time diameter is at most
  $1+\sqrt{2}$ times the travel-time diameter of an optimal
  axis-aligned speed-$v$ highway cross for $S$.
\end{theorem}

\begin{proof}
  According to Theorem~\ref{thm:cross-infinite} the median highways
  can be computed in $O(n \log n)$ time.  According to
  Lemma~\ref{lem:median} they yield a factor-$(2+1/v)$ approximation
  for the optimal travel-time diameter.  Note that the approximation
  factor tends to~3 when the speed goes to~1.  Clearly \emph{not}
  building a highway cross is a factor-$v$ approximation.  Balancing
  out the two terms yields $\min\{ 2+1/v, v\} \leq 1+\sqrt{2}$.
\end{proof}

The following lemma will allow us to do better.

\begin{lemma}\label{lem:twoapprox}
  Let $s$ be any point in $S$.  Let $H_s$ be the highway cross that
  minimizes the maximum travel time to $s$.  The travel-time diameter
  of $H_s$ is at most twice the travel-time diameter $\deltaopt$ of an
  optimal axis-aligned speed-$v$ highway cross for $S$.
\end{lemma}

\begin{proof}
  Let $\{p,q\}$ be a pair of points in $S$ and let $s' \in S$ be a
  point of maximum travel-time distance from $s$ given $H_s$.  We
  denote by $d_s$ the metric induced by $H_s$ and by \dopt\ the metric
  induced by the optimal highway cross.  Then $d_s(s,p) \leq d_s(s,s')
  \leq \dopt(s,s') \leq \deltaopt$ and, by symmetry, $d_s(s,q) \leq
  \deltaopt$.  This yields $d_s(p,q) \leq d_s(p,s) + d_s(s,q) \leq 2
  \deltaopt$.
\end{proof}

Note that $H_s$ is usually \emph{not} centered at $s$ (consider, for
instance, the set $S=\{(0,1),(1,0)\}$ whose optimal highway cross is
centered at the origin).

Based on the constant-factor approximation from
Theorem~\ref{thm:constbootstrap} we can use binary search and the
decision procedure of Theorem~\ref{thm:decision} to get the following.

\begin{theorem}
  Given a set $S$ of $n$ points in the plane, we can compute in
  $O(\log (1/\eps) \alpha(n) n \log n)$ time a
  $(2+\eps)$-approximation for the optimal axis-aligned speed-$v$
  highway cross for~$S$.
\end{theorem}

\begin{proof} 
  Let $s$ be any point in $S$ and $H_s$ be the highway cross that
  minimizes the maximum travel time to $s$.

  According to Theorem~\ref{lem:twoapprox} the travel-time diameter
  $\delta_{s}$ given $H_s$ is at most twice the travel-time diameter
  $\deltaopt$ given an optimal axis-aligned speed-$v$ highway cross
  for $S$, that is, $\delta_{s} \leq 2\deltaopt$, so a
  $(1+\eps/2)$-approximation $\delta_{\eps}$ to $\delta_{s}$ is a
  $(2+\eps)$-approximation for $\deltaopt$.

  We now describe how to compute a $(1+\eps)$-approximation for
  $\delta_{s}$ by binary search.  Recall that the median highways
  yield a travel-time diameter of $\deltamed \leq (1+\sqrt{2})
  \deltaopt \leq 3 \deltaopt \leq 3 \delta_{s} $ of $\delta_{s}$, see
  Theorem~\ref{thm:constbootstrap}.  The median highways can be
  computed in $O(n \log n)$ time according to
  Theorem~\ref{thm:cross-infinite}.  

  Now we conceptually subdivide the interval $I=[0,2\deltamed]$ into
  at most $N=6/\eps$ pieces of length $\deltamed \cdot \eps/3$, and
  denote the increasing sequence of interval endpoints by
  $\Delta=(\delta_{1}, \dots, \delta_{N})$. Since $\delta_{s} \leq
  2\deltaopt \leq 2 \deltamed$, we know that $\delta_s$ lies in $I$.
  Hence there is an index $i \in \{1, \dots, N\}$ such that $\delta_{i}
  < \delta_{s} \leq \delta_{i+1}$.

  Setting $\delta_{\eps} = \delta_{i+1}$ we find that 
  $\delta_{s} \geq \delta_{i} = \delta_{\eps} - \deltamed \cdot \eps/3$.
  This yields $\delta_{\eps} \leq \delta_{s} + \deltamed \cdot \eps/3
  \leq \delta_{s} + 3\delta_{s} \cdot \eps/3 \leq (1+ \eps)
  \delta_{s}$.  Thus $\delta_{\eps}$ is indeed a
  $(1+\eps)$-approximation of $\delta_{s}$.

  For a given $\delta > 0$ we run the decision algorithm of
  Theorem~\ref{thm:decision}, using the set of $n-1$ pairs $P = \{(s,
  q) \mid q\in S \setminus \{s\}\}$.  Each such test takes
  $O(\alpha(n) n \log n)$ time.  Using $O(\log (1/\eps))$ calls to
  this decision procedure, we can determine $\delta_{\eps}$ by binary
  search on $\Delta$. We return the highway cross computed by the
  decision procedure for the largest $\delta_{i} \leq \delta_{s}$.
\end{proof}

If we are willing to invest more time, we can even get a
$(1+\eps)$-approximation of the optimal travel-time diameter \deltaopt.

\begin{theorem}
  Given a set $S$ of $n$ points in the plane, we can compute in
  $O(\log (1/\eps) \alpha(n) n^2 \log n)$ time a
  $(1+\eps)$-approximation for the travel-time diameter of $S$ under
  the optimal axis-aligned speed-$v$ highway cross.
\end{theorem}

\begin{proof}
  We again first compute the median highways to get an upper bound
  \deltamed\ for the optimal travel-time diameter \deltaopt\ and then
  do binary search.  We can now use the interval $[0, \deltamed]$,
  which contains \deltaopt.  We stop when the interval size is
  sufficiently small, that is, at most $\deltamed \cdot \eps/3$.  This
  time we use the decision algorithm of
  Theorem~\ref{thm:decision} with the set $P$ of all $n \choose 2$
  pairs of points in~$S$.
\end{proof}

\section{Concluding remarks}

There are many ways how this problem can be extended.  First, can we
compute an optimal highway with arbitrary orientation under the
Euclidean metric in $o(n^2)$ (worst-case) time?  Second, consider
highways with different speeds, different slopes, or bounded lengths.
Third, suppose an existing network of (axis-parallel) highways and a
real $\ell > 0$ is given.  Where to place a new (axis-parallel)
highway segment of length $\ell$ in order to minimize the travel-time
diameter of the resulting network?

\bibliographystyle{geobook}
\bibliography{abbrv,highway}

\begin{thebibliography}{10}

\bibitem{ahiklmps-pptml1in-01}
M.~Abellanas, F.~Hurtado, C.~Icking, R.~Klein, E.~Langetepe, L.~Ma, B.~{Palop
  del R{\'\i}o}, and V.~S{\'a}cristan.
\newblock Proximity problems for time metrics induced by the {$L_1$} metric and
  isothetic networks.
\newblock In {\em Actas de los IX Encuentros de Geometr{\'i}a Computacional},
  pages 175--182, Girona, 2001.

\bibitem{ahiklmps-vdsnh-03}
M.~Abellanas, F.~Hurtado, C.~Icking, R.~Klein, E.~Langetepe, L.~Ma, B.~{Palop
  del R{\'\i}o}, and V.~S{\'a}cristan.
\newblock {Voronoi} diagram for services neighboring a highway.
\newblock {\em Inform. Process. Lett.}, 86:283--288, 2003.

\bibitem{aap-qpsscvd-04}
O.~Aichholzer, F.~Aurenhammer, and B.~{Palop del R{\'\i}o}.
\newblock Quickest paths, straight skeletons, and the city {Voronoi} diagram.
\newblock {\em Discrete Comput. Geom.}, 31:17--35, 2004.

\bibitem{bc-spvdtngp-05}
S.~W. Bae and K.-Y. Chwa.
\newblock Shortest paths and {Voronoi} diagrams with transportation networks
  under general distances.
\newblock In {\em Proc. 16th Annu. Internat. Sympos. Algorithms Comput.
  (ISAAC'05)}. {\em Lecture Notes Comput. Sci.}, vol.~3827, pages 1007--1018.
  Springer-Verlag, 2005.

\bibitem{bc-vdtnep-06}
S.~W. Bae and K.-Y. Chwa.
\newblock Voronoi diagrams for a transportation network on the {Euclidean}
  plane.
\newblock {\em Internat. J. Comput. Geom. Appl.}, 16:117--144, 2006.

\bibitem{bkc-occvd-06}
S.~W. Bae, J.-H. Kim, and K.-Y. Chwa.
\newblock Optimal construction of the city {Voronoi} diagram.
\newblock In {\em Proc. 17th Annu. Internat. Sympos. Algorithms Comput.
  (ISAAC'06)}. {\em Lecture Notes Comput. Sci.}, vol.~4288, pages 183--192.
  Springer-Verlag, 2006.

\bibitem{cl-mgflp-06}
J.~Cardinal and S.~Langerman.
\newblock Min-max-min geometric facility location problems.
\newblock In {\em Proc. of the European Workshop on Computational Geometry
  (EWCG'06)}, pages 149--152, Delphi, March 2006.

\bibitem{c-garot-99}
T.~M. Chan.
\newblock Geometric applications of a randomized optimization technique.
\newblock {\em Discrete Comput. Geom.}, 22:547--567, 1999.

\bibitem{egs-ueplf-89}
H.~Edelsbrunner, L.~J. Guibas, and M.~Sharir.
\newblock The upper envelope of piecewise linear functions: algorithms and
  applications.
\newblock {\em Discrete Comput. Geom.}, 4:311--336, 1989.

\bibitem{fj-gsrsm-84}
G.~N. Frederickson and D.~B. Johnson.
\newblock Generalized selection and ranking: sorted matrices.
\newblock {\em SIAM J. Comput.}, 13:14--30, 1984.

\bibitem{gsw-ccvdf-07}
R.~G{\"o}rke, C.-S. Shin, and A.~Wolff.
\newblock Constructing the city {Voronoi} diagram faster.
\newblock {\em Internat. J. Comput. Geom. Appl.}, 2007.
\newblock To appear.

\bibitem{sa-dssga-95}
M.~Sharir and P.~K. Agarwal.
\newblock {\em Davenport-Schinzel Sequences and Their Geometric Applications}.
\newblock Cambridge University Press, Cambridge, 1995.

\bibitem{t-sgprc-83}
G.~T. Toussaint.
\newblock Solving geometric problems with the rotating calipers.
\newblock In {\em Proc.~IEEE MELECON}, pages 1--4, Athens, Greece, 1983.

\end{thebibliography}

\end{document}